\documentclass[12pt,preprint]{aastex}
\usepackage{graphicx}    

\usepackage{amsmath}
\usepackage{epstopdf}
\usepackage{epsfig}

\newcommand\beq{\begin{equation}}
\newcommand\eeq{\end{equation}}
\newcommand\beqar{\begin{eqnarray}}
\newcommand\eeqar{\end{eqnarray}}

\newcommand\etal{{\it et al.~}}

\def\ga{\mathrel{\raise.3ex\hbox{$>$\kern-.75em\lower1ex\hbox{$\sim$}}}}
\def\la{\mathrel{\raise.3ex\hbox{$<$\kern-.75em\lower1ex\hbox{$\sim$}}}}

\newcommand\iun{\mbox{${\rm \, photons \,\, cm^{-2} \, s^{-1} \, sr^{-1}}\ $}}

\newcommand\fun{\mbox{${\rm \, photons \,\, cm^{-2} \, s^{-1} \, GeV^{-1} \,sr^{-1}}$}}

\begin{document}

\title{Diffuse Gamma Rays from the Galactic Plane: \\ 
Probing the ``GeV Excess'' and Identifying the ``TeV Excess''}

\author{Tijana Prodanovi\'{c}, Brian D. Fields\altaffilmark{1}
\altaffiltext{1}{also Department of Physics, University of Illinois}}
\affil{Department of Astronomy, University of Illinois, Urbana,
Illinois 61801, USA}
%
%\and
%
\author{John F. Beacom\altaffilmark{2}
\altaffiltext{2}{also Department of Astronomy, The Ohio State University}}
\affil{Department of Physics, The Ohio State University, Columbus,
Ohio 43210, USA}

%%%%%%%%%%%%%%%%%%%%%%%%%%%%%%%%%%%%%%%%%%%%%%%%%
%%%%%%%%%%%%%%%%%%%%%%%%%%%%%%%%%%%%%%%%%%%%%%%%%

\begin{abstract}
Pion decay gamma rays have long been recognized as a unique signature
of hadronic cosmic rays and their interactions with the interstellar
medium. We present a model-independent way of constraining this signal
with observations of the Galactic Plane in diffuse gamma rays.  We
combine detections by the EGRET instrument at GeV energies and the
Milagro \v{C}erenkov detector at TeV energies with upper limits from
KASCADE and CASA-MIA ground arrays at PeV energies.  Such a long
``lever arm'', spanning at least six orders of magnitude in energy,
reveals a ``TeV excess'' in the diffuse Galactic Plane gamma-ray
spectrum.  While the origin of this excess is unknown, it likely
implies also enhanced TeV neutrino fluxes, significantly improving the
prospects for their detection.  We show that unresolved point sources
are a possible source of the TeV excess.  In fact, the spectra of the
unidentified EGRET sources in the Milagro region must break
between $\sim 10 \rm \ GeV$ and $\sim 1 \rm \ TeV$ to avoid strongly
overshooting the Milagro measurement; this may have important
implications for cosmic-ray acceleration.
  
Finally, we use our approach to examine the recent suggestion that
dark-matter annihilation may account for the observed excess in
diffuse Galactic gamma-rays detected by EGRET at energies above 1 GeV.
Within our model-independent approach, current data cannot rule this
possibility in or out; however we point out how a long ``lever arm''
can be used to constrain the pionic gamma-ray component and in turn
limit the ``GeV excess'' and its possible sources.  Experiments such
as HESS and MAGIC, and the upcoming VERITAS and GLAST, should be able
to finally disentangle the main sources of the Galactic gamma rays.

\end{abstract}

%%%%%%%%%%%%%%%%%%%%%%%%%%%%%%%%%%%%%%%%%%%%%%%%%
%%%%%%%%%%%%%%%%%%%%%%%%%%%%%%%%%%%%%%%%%%%%%%%%%

\section{Introduction}

The Milky Way Galactic Plane has long been known to be a strong source
of diffuse gamma-ray emission \citep{Kraushaar,Fichtel}.  The
Energetic Gamma Ray Emission Telescope (EGRET) instrument on the
Compton Gamma-Ray Observatory satellite measured this emission over
the full sky and for energies in the range $0.03-10$ GeV, with
reasonably high resolution in each bin \citep[][where the data are
reported with angular bins of width 0.5 degree and with several
logarithmically-spaced bins per decade in energy]{Hunter}.  It was
expected that a very significant component of the diffuse emission
would arise from the decays of neutral pions ($\pi^0 \rightarrow
\gamma + \gamma$), arising from the collisions of hadronic cosmic rays
with the hadronic component of the interstellar medium \citep[i.e., $p
+ p \rightarrow p + p + \pi^0$;][]{Stecker70,Stecker71,Dermer}.  We
refer to these throughout the paper as ``pionic'' gamma rays, to
distinguish them from gamma rays produced by leptonic
processes, e.g., the inverse-Compton upscattering of ambient photons
by very high-energy electrons.

The spatial variation of the pionic component of the diffuse Galactic
gamma-ray emission should track the column density of the interstellar
medium.  However, since other sources of gamma rays also depend,
though in more complicated ways, on the imprecisely-known distribution
of interstellar matter and radiation, it is difficult to extract the
pionic component by its spatial dependence alone.  However, the energy
spectrum of the pionic gamma rays has a characteristic shape, which
follows from the kinematics of the boosted pion decays and the
convolution with the hadronic cosmic ray spectrum.  This shape is
symmetrical (when plotted as $\log dN/dE$ vs $\log E$) about the
``pion bump'' peak at $m_{\pi^0}/2 \simeq 0.07$ GeV, with long
power-law tails to higher and lower energies \citep{Stecker71}.  Since
no strong evidence of this bump was seen, \citet{Pionic} estimated the
maximum pionic fraction of the diffuse Galactic gamma-ray emission to
be $\sim 50\%$.  This is supported by the very detailed and
comprehensive study of the Galactic gamma-ray emission by \citet{SMR}.
That study indicates that a key second feature of pionic gamma rays is
that at high energies (certainly by $\sim 1$ TeV) they should dominate
the total emission and their slope will follow that of the hadronic
cosmic rays.  (The emission at energies below the bump is expected to
be subdominant to the leptonic components.)  In the GeV energy
range, a significant component of the observed EGRET data is not well
explained, and this discrepancy, which is observed in all sky
directions, is known as the ``GeV excess" \citep{Hunter}.

In this paper, we consider constraints on the pionic gamma rays from
experiments operating at much higher energies than EGRET.  There are
upper limits on the total gamma-ray emission near both TeV ($= 10^3$
GeV) and PeV ($= 10^6$ GeV) energies.  Depending on assumptions about
the slope of the hadronic cosmic ray spectrum, these place at least
somewhat restrictive limits on the pionic gamma-ray emission.
However, the most exciting recent development is the first positive
{\em detection} of diffuse gamma-ray emission from the Galactic Plane,
by the Milagro Collaboration \citep{Milagro05}.  They find
$\phi_\gamma (> 3.5 \ \rm TeV) = (6.8 \pm 1.5 \pm 2.2) \times
10^{-11}$ \iun in the Galactic Plane region of longitude $\ell \in
(40^{\circ}, 100^{\circ})$ and latitude $|b|<5^{\circ}$.  The basic
question of the present paper is ``What is the origin of the
(apparently) diffuse flux observed by Milagro?''.  The Milagro
Collaboration argued that their result is consistent with being purely
pionic in origin, though they note that some of the flux may arise
from unresolved point or extended sources \citep{Milagro05}.  As we
will argue in steps of increasing detail, their result appears to be
{\em too large} to be purely pionic, and thus seems to indicate a new
mystery of Galactic gamma rays, which we will call the ``TeV excess."
Our GeV--TeV--PeV overview perspective is shown in Fig.~\ref{fig:all}.
In brief, our arguments are as follows:
\begin{enumerate}
\item We can simply extrapolate the last EGRET points to higher
energies, and the Milagro result should not exceed this trend -- and
while it does not, it could not be any larger.  This is shown in
Fig.~\ref{fig:all} as the solid line.\footnote{ The Milagro
Collaboration extrapolated the {\em integral} energy spectrum, while
we use the {\em differential} energy spectrum (though adopting the
same spectral index).  While these procedures are in principle
equivalent, the smoothness of the integral spectrum comes at the cost
of correlations between the points.}  This alone indicates that a
strong inverse Compton component at high energies is disfavored
\citep{Milagro05}, in agreement with considerations at lower energies
\citep{SMR}.  In the PeV range, this extrapolation appears to be at
best barely allowed, and is possibly excluded, depending on the choice
of the hadronic cosmic ray spectrum.
\item A more sophisticated approach is to only extrapolate the pionic
component to high energies, where it should dominate.  We first
consider a pionic component of maximal normalization (while this is
unrealistically high, it is in fact {\it lower} than the normalization
obtained if the GeV EGRET data is effectively assumed to be purely
pionic, as above).  We allow two choices for the hadronic cosmic ray
energy spectrum slope at high energies, as shown by the dashed lines
in Fig.~\ref{fig:all}.  The first, with index $\alpha = 2.61$ (for
$d\phi/dE \sim E^{-\alpha}$), is motivated by the slope chosen by the
Milagro Collaboration, which provides a good single-parameter fit
joining the highest-energy EGRET points to the Milagro point.  We
argue below that the physics of pionic emission suggests that this is
an unrealistically shallow spectrum if the observed GeV--TeV signal is
indeed pionic.  When we instead choose $\alpha = 2.75$, in accordance
with the locally observed cosmic rays \citep{JACEE}, we find that the
Milagro measurement is 5 times {\em larger} than the {\em maximal}
pionic flux allowed at 3.5 TeV.  The PeV limits are on the verge of
ruling out (or detecting!) the pionic signal, regardless of the choice
of $\alpha$.  In addition, when the PeV limits are derived using local
cosmic ray spectrum, this rules out the continuation of GeV-TeV
$\alpha = 2.61$ gamma-ray power law to PeV energies.
\item More realistically, the normalization of the pionic component
should be even lower, at most $\sim 50\%$ of maximal in the
``optimized'' model of \citet{SMR} designed to minimize the GeV
excess.  On the other hand, the ``conventional'' model of \citet{SMR},
which uses the locally observed cosmic-ray spectrum and normalization,
comes somewhat closer to the EGRET data near the pionic maximum at
$m_\pi/2$, but leaves the GeV excess (thus motivating the non-standard
optimized model).  The results from the conventional model appear as
the dotted line in Fig.~\ref{fig:all}.  As noted by \citet{deBoer},
the GeV excess of the conventional model allows room for a large
component of gamma rays from dark matter annihilation products
(including pions, though we reserve the word ``pionic" to refer to
pions produced by cosmic ray collisions with the interstellar medium).
These gamma rays from dark matter are claimed to help ameliorate the
GeV excess (note that their spectrum abruptly ends below the dark
matter mass of 50--100 GeV).  We point out here that this
interpretation {\em increases} the TeV excess, making the Milagro
measurement about 10 times larger than the pionic component.
\end{enumerate}

Thus, taking a realistic normalization and slope for the pionic
component, we find that the Milagro measurement seems to indicate a
TeV excess, which would be even more interesting than their conclusion
that their result may be consistent with being purely pionic.  Our
arguments are supported by the gamma-ray flux limits at PeV energies.
The diffuse gamma-ray data is summarized in \S\ref{sect:data}.  In
\S\ref{sect:analysis} we analyze the consistency of the data with
diffuse pionic emission, and explore the possibility of unresolved
sources contributing significantly to the Milagro measurement.  We go
on in \S\ref{sect:discuss} to show how the framework of the
GeV--TeV--PeV Galactic gamma-ray emission can be tested in detail.  We
conclude in \S\ref{sect:conclude} with an observational strategy which
uses present and upcoming gamma-ray experiments to disentangle the
nature of diffuse Galactic gamma-ray sources, both pionic and
otherwise.

The resolution of the outstanding issues has important implications
for more than just the pionic gamma rays, and will shed new light on
Galactic cosmic rays in numerous ways: it will probably finally
detect, and at least strongly constrain, the presence and interactions
of hadronic cosmic rays throughout the Galactic interstellar medium;
it will constrain the origin, source distribution, and spectra of both
hadronic and leptonic cosmic rays; and it will thereby sharpen our
account of the Galactic cosmic ray energy budget and thus the
efficiency of cosmic ray accelerators.  Furthermore, a detailed and
quantitative understanding of astrophysical sources of diffuse
Galactic gamma-rays will greatly clarify the existence and nature
of any other Galactic sources, such as dark matter.  And finally, a
good understanding of Galactic gamma-rays will allow for this
foreground to be better subtracted to obtain the diffuse extragalactic
gamma-ray background.

%%%%%%%%%%%%%%%%%%%%%%%%%%%%%%%%%%%%%%%%%%%%%%%%%
%%%%%%%%%%%%%%%%%%%%%%%%%%%%%%%%%%%%%%%%%%%%%%%%%

\section{High-Energy Gamma-Ray Data}
\label{sect:data}

The Milagro Gamma-Ray Observatory is a ground-based water \v{C}erenkov
detector in New Mexico that collects air-shower particles created when
high-energy particles interact in the atmosphere; showers initiated by
gamma-rays and hadrons can be statistically distinguished by how they
register in the detector \citep{Milagro03, Milagro04, Milagro05}.  The
Milagro Collaboration recently reported a diffuse flux $\phi_\gamma
(>3.5 \ \rm TeV) = (6.8 \pm 1.5 \pm 2.2) \times 10^{-11}$ \iun of
gamma rays from the Galactic Plane region $\ell \in (40^{\circ},
100^{\circ})$ and latitude $|b|<5^{\circ}$ \citep{Milagro05}.  Note
that this emission is integrated over both higher energies and also
the entire angular region, where no resolved sources were detected
\citep{Milagro05}.  In fact, to obtain the Galactic Plane diffuse
emission Milagro did not directly measure the gamma-ray flux, but
rather the ratio of electromagnetic to hadronic showers.  Furthermore,
their measurement was made by subtracting the off-source and on-source
(Galactic Plane) fluxes, in order to cancel the isotropic cosmic-ray
component; this also cancels the extragalactic gamma-ray background,
which is at the otherwise relatively high level $\sim 10^{-6}$ GeV
cm$^{-2}$ s$^{-1}$ sr$^{-1}$ (compare in Fig.~\ref{fig:all}).  An
independent measurement of the hadronic cosmic-ray flux was then taken
to derive the gamma-ray flux, and the result also depends on the
assumed spectral indices of each species.  We note that for the
hadronic cosmic rays, Milagro adopts the conventional observed value
$\alpha = 2.75$.

The Milagro Collaboration reports that their result is consistent with
the diffuse emission extrapolated from EGRET, assuming a spectral
index $\alpha = 2.61$, which is taken from the last four points of the
EGRET integral spectrum \citep{Milagro05}.  This single-parameter fit
provides a good description of these data.  (By extrapolating from the
EGRET differential spectrum, our Fig.~\ref{fig:all} highlights the
uncertainty in this procedure, that is, it demonstrates how a small
change in the assumed spectral index can be important over a large
energy range.)  The apparent success of a single power law over this
large energy baseline is very suggestive that the emission at these
energies is dominated by a single source.  In particular, given the
understanding of how the various components of the diffuse emission
change with energy from \citet{SMR}, one sees that this effectively
assumes that all of the EGRET GeV data is pionic.  However, the
Milagro Collaboration is careful to note that their result could have
a contribution from unresolved point or extended sources
\citep{Milagro05, Nemethy}.
 
This first detection of diffuse emission at TeV energies invites a
detailed comparison with other data.  In our analysis, we will start
with the assumption that Milagro detection corresponds to truly
diffuse pionic emission, and then investigate the validity and
consequences of this.

The EGRET data covers the range $0.03-10 \rm \ GeV$ and is publicly
available from the NASA archives \footnote[1]{EGRET data archive:
http://cossc.gsfc.nasa.gov/docs/cgro/egret/} in the form of integral
gamma-ray fluxes (in a given energy bin) at a given galactic
coordinate where the coordinate step is $0.5^{\circ}$.  We have taken
those data points that fall into the Milagro region $\ell \in
(40^{\circ}, 100^{\circ}),|b|<5^{\circ}$, and averaged them for each
energy bin. Finally, we have determined the EGRET gamma-ray flux at
the mean energy for each bin, where the underlying assumption is that
the flux is energy-independent over the width of a bin.  This is
presented in Fig.~\ref{fig:all} with red data points.  Following
\citet{SMR}, we took fixed fractional uncertainties of $15\%$ on the
fluxes (since these are predominantly systematic in nature, they do
not change when the field of view changes).  Below, we additionally
consider the EGRET sources detected in this region, taking their
spectra from the Third EGRET Catalog \citep{Hartman}.

We also consider the upper limits on gamma-ray fluxes from other
high-energy experiments.  Although these experiments did not observe
exactly the same region of the Galactic Plane as Milagro, we argue
that their results can be put on a common footing.  Especially at and
above 1 TeV, it is expected that the diffuse Galactic emission is
purely pionic, and hence scales with the column density \citep{SMR}.
Then fluxes from different regions of the Galactic Plane, if corrected
for differences in column density, can be made {\em physically
equivalent}, even if they are {\em geographically distinct}.  This
depends on the common assumption that there are no significant
variations in the hadronic cosmic ray fluxes and spectra as a function
of position in the Galactic Plane \citep[e.g.,][but see also
\citet{smrdd}]{sm96}.

To correct for the differences in column density in different regions
of the Galactic Plane, we take a simplistic approach and scale from
the EGRET diffuse flux at lower energies (even though it is not purely
pionic at those energies, this should be a reasonable approach for the
relative variations in expected intensity).  We calculate the region
correction factor by comparing the EGRET diffuse gamma-ray flux
averaged over the Milagro region with the one averaged over the region
observed by a given experiment.  We find that our correction factors
do not vary much with energy.  Table 1 summarizes the input data and
the region correction factors $f_{\rm rc}$. Here, $f_{\rm rc}=F_{\rm
EG,reg}/F_{\rm EG,Milagro}$ where $F_{\rm EG,reg}$ and $F_{\rm
EG,Milagro}$ are the diffuse gamma-ray flux observed by EGRET and
averaged over a given Galactic region and the region observed by
Milagro, respectively.

For energies near 1 TeV, we show in Fig.~\ref{fig:all} the equivalent
upper limits on the diffuse Galactic gamma-ray emission from the
Whipple \citep{Whipple}, HEGRA \citep{HEGRA} and Tibet-$\rm II/III$
\citep{tibet} experiments.  For energies near 1 PeV, we also show the
similar upper limits from the CASA-MIA \citep{CASA-MIA} and KASCADE
\citep{KASCADE} experiments.

The diffuse gamma-ray limits reported have an underlying assumption of
a spectral index.  We present each observational limit as originally
reported with their assumed spectral index.  For CASA-MIA, only the
ratio of gamma-ray to hadronic integrated fluxes was reported in
\citet{CASA-MIA}, and we take the spectral index given by
\citet{CASA-MIA_spec}.  We have to note here that there is a strong
dependence of CASA-MIA limits on the assumed spectral index.  This
point is emphasized in Fig.~\ref{fig:zoom} where we plot the CASA-MIA
limits for their measured spectral index $\alpha=2.66$
\citep{CASA-MIA_spec}, and also for the steeper spectral index of
$\alpha=2.80$ reported by JACEE \citep{JACEE}.  On the other hand, the
KASCADE limits \citep{KASCADE} do not depend on the assumption of the
spectral index \citep{Schatz}.

%%%%%%%%%%%%%%%%%%%%%%%%%%
\begin{table}[h]
\begin{center}
\caption{Diffuse gamma-ray observations used in this paper.  The flux
limits quoted by the various experiments are divided by $f_{rc}$ before
being shown in our Fig.~\ref{fig:all}; this compensates for the differences
in expectations for different regions.
\label{table:obs}}
\vspace{0.2cm}
\begin{tabular}{cccccc}
\tableline\tableline
 & \multicolumn{2}{c}{Observation Region} & Region & Spectral 
& Confidence \\
Experiment& $\ell$ range & $|b|$ range & Correction  $f_{\rm rc}$ & 
Index $\alpha $ & Limit\\
\tableline\tableline
Milagro & $(40^{\circ}, 100^{\circ})$ & $<5^{\circ}$ & $\equiv 1$ & 2.61 &
99 \% \\
\tableline%\tableline
Whipple  & $(38.5^{\circ}, 41.5^{\circ})$ & $<2^{\circ}$ & 1.6 & 2.4 &
99.9 \% \\
%\tableline
HEGRA  & $(38^{\circ}, 43^{\circ})$ & $<2^{\circ}$ & 1.6 & 2.6 & 99 \% \\
%\tableline
Tibet  II, III& $(20^{\circ}, 55^{\circ})$ & $<2^{\circ}$ & 1.6 & 2.5 &
 99 \% \\
%\tableline
CASA-MIA & $(50^{\circ}, 200^{\circ})$ & $<5^{\circ}$ & 0.7 & 2.66 & 90 \%  \\
%\tableline
KASCADE & R.A. $\in (0^{\circ}, 360^{\circ})$ & $\delta \in (14^{\circ},84^{\circ})$ & 0.2 & independent & 90 \%\\
\tableline\tableline
\end{tabular}
\end{center}
\end{table}

%%%%%%%%%%%%%%%%%%%%%%%%%%

%%%%%%%%%%%%%%%%%%%%%%%%%%%%%%%%%%%%%%%%%%%%%%%%%
%%%%%%%%%%%%%%%%%%%%%%%%%%%%%%%%%%%%%%%%%%%%%%%%%

\section{Analysis of the Data}
\label{sect:analysis}

%%%%%%%%%%%%%%%%%%%%%%%%%%%%%%%%%%%%%%%%%%%%%%%%%

\subsection{Diffuse Components}

The spectrum of gamma rays that originate from the decay of neutral
pions created in hadronic cosmic-ray interactions with ambient protons
($p_{\rm cr} + p_{\rm ism} \rightarrow p + p + \pi^0$, followed by
$\pi^0 \rightarrow \gamma + \gamma$) is given in a useful fit by
\citet{Ensslin}, essentially following the earlier model by
\citet{Dermer}.  The symmetry of the two-gamma emission guarantees
that the photon spectrum is symmetric about a peak at $m_\pi/2$ when
plotted with a log scale \citep[e.g.,][]{Stecker70, Stecker71}.
Furthermore, the asymptotic logarithmic slope (i.e., the spectral
index $\alpha$) at high energies is the same as that of the primary
cosmic rays \citep[e.g.,][]{Dermer, Mori, Kamae}.  Thus, the pionic
spectral shape is determined by a single parameter, the cosmic-ray
spectral index.  However, we note that there are still uncertainties
in this pionic source function; see the discussions in e.g.,
\citet{Blattnig} and \citet{Kamae}.  At the present level of analysis,
the uncertainties in the astrophysical inputs, particularly the
Galactic cosmic-ray spectrum, are larger.\footnote{For example,
\citet{Kamae} finds that diffractive effects could change the
gamma-ray index by about +0.05 units; this is about the level of the
uncertainty in the measured local cosmic-ray spectrum, but much
smaller than the index shift ($\ge 0.2$ units) needed to reconcile the
EGRET and Milagro data with the pionic signal expected from
cosmic rays.}  \citet{Pionic} have shown that the lack of a strong
pionic feature at $m_\pi/2$ in the diffuse Galactic gamma-ray data can
be used to place a model-independent (i.e., flux-independent) upper
limit on the pionic component of $\sim 50\%$.

For better comparison to other data, we assume a spectral index and
convert the Milagro energy-integrated flux into a differential flux,
also evaluated at 3.5 TeV.  If we adopt the Milagro best-fit gamma-ray
index of $\alpha = 2.61$, we find a gamma flux of $d\phi/dE = (3.1 \pm
1.2 ) \times 10^{-14} \fun$.  This point is shown in
Fig.~\ref{fig:all} as a blue triangle.  We note here that if the
integral flux reported by Milagro is recalculated for a more realistic
spectral index of $\alpha=2.75$ then the variation of the flux is just 6\%,
which is much smaller than reported uncertainty.

Even when the pionic component is maximized \citep{Pionic}, it fails
to explain the Milagro result.  To appreciate this mismatch, it is
important to recall that the physics of the pionic signal demands that
above the pion bump, the pionic spectrum is characterized by a single
spectral index which is the same as that of the cosmic rays.  Thus, if
the high-energy EGRET and Milagro points are due to pionic emission,
their spectral index must reflect the underlying cosmic ray index
along the line of sight.  If we adopt the best-fit EGRET/Milagro index
$\alpha = 2.61$ as reflecting the average Galactic cosmic ray spectrum
towards the Milagro region, the resulting pionic flux at the Milagro
energy is 66\% of the observed result.  For the locally-measured
cosmic-ray spectral index of $\alpha = 2.75$, the maximal allowed
pionic contribution drops to just 19\% of the Milagro flux.  Note
however that due to large uncertainties in Milagro measurement, the
maximal fraction that the pionic gamma-ray component can account for
in this case, can be at most about $30 \%$.  Were we to raise the
pionic prediction to meet the Milagro and high-energy EGRET signals,
the result would overshoot the EGRET signal below 1 GeV.

This result on the pionic normalization, supported by the more
detailed work of \citet{SMR}, indicates that it is not realistic to
simply extrapolate the EGRET data into the TeV range, where the pionic
component should be dominant.  At the very least, the non-pionic
components of the GeV data should be subtracted first.  Also, as shown
by the solid line in Fig.~\ref{fig:all}, when the EGRET data are
further extrapolated into the PeV range, the expectations are right on
the edge of upper limits from the CASA-MIA and KASCADE experiments.
The upper dashed line in Fig.~\ref{fig:all} shows a line of the same
EGRET/Milagro best-fit spectral index ($\alpha = 2.61$), with a
maximal pionic normalization.  Besides being $\sim 2$ times larger
than favored at low energies, this curve still falls below the Milagro
point (with a more realistic normalization, it would fall more
significantly below).

While the spectral index fit of $\alpha = 2.61$ is quite
suggestive for connecting the EGRET and Milagro observations, it is
not consistent with local observations of the hadronic cosmic rays,
which instead suggest $\alpha = 2.75$.  Over the long lever arm of
$\sim 1$ GeV to $\sim 1$ TeV, this makes a significant difference.
Cosmic-ray experiments such as JACEE fit their measured cosmic-ray
spectra with $\alpha = 2.80 \pm 0.04$ \citep{JACEE}.  In our analysis
we will adopt $\alpha=2.75$ as a more conventional, locally measured
value, consistent with our previous work.  In this case, we find that,
even for a maximized pionic normalization, the pionic flux at 3.5 TeV
is 5 times smaller than the Milagro measurement.  For a pionic
normalization as low as assumed by \citet{deBoer}, the pionic flux at
3.5 TeV is about 10 times smaller than the Milagro measurement.  In
any case, the joint demands of using a realistic cosmic ray spectrum
and not exceeding the maximal pionic normalization mean that the
expectations fall well below the Milagro observation.  We therefore
call this problem the ``TeV excess."

Pushing beyond the TeV range to PeV energies further constrains both
the TeV and GeV excesses.  In Fig.~\ref{fig:all}, we see that the
upper limits reported by CASA-MIA \citep{CASA-MIA} and KASCADE
\citep{KASCADE} appear to already rule out the simple single-power-law
extrapolation from GeV energies upward.  Indeed, the published PeV
limits barely permit the maximal pionic emission allowed at an index
of at the level of the EGRET/Milagro $\alpha = 2.61$ fit.  Thus the
PeV data already play a useful role in limiting the level of pionic
emission and thus strengthening the case for a non-pionic TeV excess
seen by Milagro.  Indeed, it is clear that there is no source which
can have a single power law spectrum which lies beneath the GeV data
and matches the TeV signal, without running afoul of the PeV
constraints.

Moreover, as noted above, the PeV data from CASA-MIA were obtained
from a gamma-to-hadron shower ratio in concert with an assumed
cosmic-ray spectral index of $\alpha = 2.66$.  In Fig.~\ref{fig:zoom},
we zoom into the TeV--PeV region to show the effect of choosing the
steeper spectral index $\alpha = 2.80$ measured by \citet{JACEE}.
Note that because only the ratio of integral fluxes is given, the
assumption of a different spectral index also results in a different
normalization need to calculate gamma-ray flux.  We then see that the
limits can become much stronger in absolute terms.  The pionic
constraints remain similar, as both the data and predictions move
together.  On the other hand, the tighter absolute limits now exclude
a continuation of the GeV-TeV $\alpha = 2.61$ power-law fit to PeV
energies.

\citet{Moskalenko} have recently shown that the attenuation of gamma
rays by the interstellar radiation field ($\gamma + \gamma \rightarrow
e^+ + e^-$) can be significant for energies $\gtrsim 10$ TeV and
sightlines near the Galactic Center.  This effect would be most
prominent around $100 \ \rm TeV$. However, at few hundred TeV
attenuation by the CMB takes over and dominates at PeV energies
\citep{Moskalenko}.  As the sensitivity and impact of the PeV data
improves, it will be necessary to take these effects into account.  In
addition, the decreasing flux and heavier composition beyond the
cosmic ray knee will also reduce the expected gamma fluxes.

The presence of a TeV excess must be viewed in the light of the
well-known GeV excess and its possible explanations.  Inverse Compton
scattering makes a significant contribution at GeV energies, but in
the TeV regime it declines rapidly, and is much smaller than the
pionic gamma-ray flux \citep{SMR}.  In the \citet{deBoer} proposed
scenario, the GeV excess originates from the annihilation of dark
matter particles with mass $\simeq 100$ GeV.  In this case the
dark-matter gamma-ray signal will have a sharp cutoff at the dark
matter mass, and again cannot contribute as significantly at TeV
energies.  (And since we are discussing the Galactic Plane, well away
from the center, the contribution of any dark matter component should
be greatly reduced.)  In order to explain the TeV excess, we require a
component which is subdominant at GeV energies, important at TeV
energies, and vanishing again at PeV energies.  This might arise from
unresolved sources with hard ($\alpha \simeq 2$) spectra, cutting off
before the PeV range, and we turn to this possibility next.

%%%%%%%%%%%%%%%%%%%%%%%%%%%%%%%%%%%%%%%%%%%%%%%%%

\subsection{Unresolved Sources}

It is possible that unresolved point or extended sources contributed
to the total gamma ray flux measured by Milagro \citep{Milagro05,
Nemethy}.  While Milagro did not find any resolved sources in this
region of the Galactic Plane, there are ten unidentified gamma-ray
point sources in this region given in the Third EGRET Catalog
\citep{Hartman}.  (It is worth noting that the definition of these as
point sources depends on the degree-scale angular resolution of EGRET;
future experiments should be able to measure the energy spectra and
angular extent of these sources much more precisely.)  The spectra of
these sources are described therein by single power law fits, which we
extrapolate to the TeV range and consider as contributions to the
Milagro diffuse measurement.

\begin{table}[h]
\begin{center}
\caption{Unidentified EGRET Point Sources in Milagro Region
\label{table:sources}}
\vspace{0.2cm}
\begin{tabular}{lrrrrr}
\tableline\tableline
3EG Catalog & \multicolumn{2}{c}{Galactic Coords} & $F(>100 \, {\rm MeV})$ 
  & Spectral Index \\
Source & $\ell$ [$^\circ$] & $b$ [$^\circ$] 
  & [$\rm 10^{-8} \, cm^{-2} \, s^{-1}$]
  & \multicolumn{1}{c}{$\gamma$} \\
\tableline
J1903+0550 & 39.52 & $-0.05$ & $62.1 \pm 8.9$ & $2.38 \pm 0.17$ \\
J1928+1733 & $52.71$ & $0.07$ & $157.0 \pm 36.9$ & $2.23 \pm 0.32$ \\
J1958+2909 & $66.23$ & $-0.16$ & $26.9 \pm 4.8$ & $1.85 \pm 0.20$ \\
J2016+3657 & $74.76$ & $0.98$ & $34.7 \pm 5.7$ & $2.09 \pm 0.11$ \\
J2020+4017 & $78.05$ & $2.08$ & $123.7 \pm 6.7$ & $2.08 \pm 0.04$ \\
J2021+3716 & $75.58$ & $0.33$ & $59.1 \pm 6.2$ & $1.86 \pm 0.10$ \\
J2022+4317 & $80.63$ & $3.62$ & $24.7 \pm 5.2$ & $2.31 \pm 0.19$ \\
J2027+3429 & $74.08$ & $-2.36$ & $25.9 \pm 4.7$ & $2.28 \pm 0.15$ \\
J2033+4118 & $80.27$ & $0.73$ & $73.0 \pm 6.7$ & $1.96 \pm 0.10$ \\
J2035+4441 & $83.17$ & $2.50$ & $29.2 \pm 5.5$ & $2.8 \pm 0.26$ \\
\tableline\tableline
\end{tabular}
\end{center}
\end{table}

In the GeV range, these objects have significantly harder spectra
($\alpha \simeq 2$) than the pionic diffuse component, so in
principle, they could become quite important at higher energies.
Additionally, we found that the combined extrapolated flux at
$E_\gamma=3.5 \ \rm TeV$ of these ten point sources, spread out over
the Milagro region, is $\sum_{i=1}^{10} F_{\rm ps}^i (E_\gamma=3.5 \rm
\ TeV) \simeq 2.5 \times 10^{-13} \fun$.  Amazingly, this is about a
factor of 10 larger than the total diffuse emission for the whole
region measured by Milagro, i.e., $F_{\rm diff} (E_\gamma = 3.5 \rm \
TeV) \approx 3.0 \times 10^{-14} \fun$.  Thus it is obvious that
indeed, unresolved point sources could contribute significantly to the
TeV excess, even taking into account the uncertainties in the
extrapolations in energy.  In fact, in order to not grossly
overproduce the measured flux, the spectra of these ten objects must
be strongly cut off before the TeV range.

Four of these ten EGRET objects have been observed at TeV energies by
the Whipple \citep{Fegan05, Buckley} and HEGRA \citep{Aharonian05}
experiments.  In Fig.~\ref{fig:sources}, we show the combined GeV and
TeV spectral information on these objects.  \citet{Aharonian05}
reported a detection by HEGRA of the source TeV J2032+4130, which, if
related to the J2033+4118 EGRET source, would mean a TeV signal that
is more than two orders of magnitude lower than what would have been
expected based on the EGRET observation.  If all of the sources were
like this, then these extrapolated unresolved sources would not be
able to explain the TeV excess.  However, in the other three cases
shown, the TeV limits are not yet strong enough to decide if these
sources are excluded from contributing significantly to the TeV
excess.  For example, even when limits for just these four sources are
used, the total flux still remains about 3 times above the Milagro
diffuse flux; and there are still the other six objects that we don't
have information about yet.

In addition, sources of comparable TeV intensity to those detected
recently by HESS \citep{HESS-Source1, HESS-Source2} could contribute
significantly to the flux in the Milagro region, if present in this
region of the Galactic Plane but not resolved; these sources may be
bright at TeV but not GeV energies.

Consequently, it is for now impossible to determine whether the
Milagro measurement arises from truly diffuse emission or unresolved
sources.  Even if the entire flux is due to unresolved sources, it is
clear that all of these ten EGRET sources will have to be cut off
before 3.5 TeV, or else the extrapolated flux could be much be too
large.  Direct observations of these ten EGRET sources in the TeV
range are thus of very high importance for further progress.

%%%%%%%%%%%%%%%%%%%%%%%%%%%%%%%%%%%%%%%%%%%%%%%%%
%%%%%%%%%%%%%%%%%%%%%%%%%%%%%%%%%%%%%%%%%%%%%%%%%

\section{Discussion}
\label{sect:discuss}

The pioneering Milagro observation \citep{Milagro05} above 3.5 TeV is
the first positive detection of a Galactic diffuse component at very
high energies.  The Milagro result becomes all the more powerful
when placed in the context of GeV gamma-ray observations by EGRET
\citep{Hunter}, and PeV upper limits by CASA-MIA \citep{CASA-MIA} and
KASCADE \citep{KASCADE}.  In particular, the combined GeV--TeV--PeV
signal is incompatible with emission from pions created by cosmic rays
with the locally measured $\alpha = 2.75$ index.  This result follows
from the physics of pion production and decay, and is independent of
any detailed Galactic model.  Moreover, pionic emission is the only
conventional source at TeV energies.  But the pionic spectral shape
and the GeV EGRET data together constrain the pionic emission to fall
below the Milagro TeV signal by at least a factor of $\sim 5$ when using a
pionic spectrum arising from cosmic rays as locally observed; even
without requiring consistency with the local cosmic-rays, the deficit
is at least a factor of $\sim 2$.  This mismatch constitutes the TeV
excess.

The TeV excess takes its place alongside the well-established GeV
excess to underscore our present state of ignorance about the sources
of diffuse Galactic gamma rays.  These data demand an explanation.
(1) We are challenged to determine the dominant sources of diffuse
Galactic gamma-rays at the highest energies, and to determine what
portion of the emission is truly diffuse, and what portion is due to
(as yet) unresolved point sources.
(2) We are still tasked to search for a pionic signature, since the
mere existence of hadronic cosmic rays and of interstellar matter
together {\em guarantee} that this flux must exist at some level in
the Galactic gamma-ray sky.
(3) Our difficulty in explaining the diffuse Galactic gamma-ray
spectrum is all the more galling given that current measurements are
consistent with a very simple spectral shape: as seen in
Fig.~\ref{fig:all}, the present GeV--TeV--PeV gamma-ray data are all
consistent with a piecewise power law having a break at a peak around
$0.8$ GeV.
It would be enormously instructive to determine whether improved
spectral resolution and energy coverage confirm this simple form or
reveal telltale features.  For now, neither the low-energy or
high-energy power law indices, nor the energy scale of the break, can
easily be understood in terms of the observed properties of local
Galactic cosmic rays.

With these broad questions at hand, we now briefly explore
astrophysical consequences of some of the possible solutions, and then
review the observational arsenal which can be brought to bear on these
problems.

%%%%%%%%%%%%%%%%%%%%%%%%%%%%%%%%%%%%%%%%%%%%%%%%%

\subsection{Point Source Spectral Break: Implications}

To account for this TeV excess we have looked into the contribution
from the unresolved sources.  More precisely, in the region of the sky
observed by Milagro we have noted ten unidentified EGRET sources. To
estimate their gamma-ray flux at $3.5 \rm \ TeV$ we have used their
spectral information as determined by EGRET and extrapolated into the
TeV energy range.  We found that the total combined gamma-ray flux of
these ten EGRET sources exceeds the diffuse Galactic Plane TeV
gamma-ray flux observed by Milagro by about an order of magnitude.
This suggests that extrapolation of the EGRET sources to TeV energies
has to fail at some point.  Four of these sources have been observed
at TeV energies, and one has been shown to break by about two orders
of magnitude.  However, until all of those point sources are surveyed
at TeV energies, we cannot say more about their possible contribution
to the TeV excess.

The possibility of a spectral break for at least some Galactic point
sources might have important implications.  If Galactic point sources
(presumably, supernova remnants) are the dominant source of Galactic
cosmic-ray protons at TeV energies then the shape of the diffuse
pionic gamma-ray spectrum should track that of individual Galactic
point sources.  That is, if there is a break or a cutoff somewhere
between $10 \rm \ GeV$ and $1 \rm \ TeV$ in gamma-ray spectra of these
sources that should carry over to spectra of cosmic rays accelerated
in them.  In that case the break in the spectrum is a measure of
maximal SNR acceleration energy.  Moreover, this would imply that
another cosmic-ray component (or reacceleration) is required to come
in before the $\sim 1000$ TeV cosmic-ray ``knee,'' contrary to
conventional models.

%%%%%%%%%%%%%%%%%%%%%%%%%%%%%%%%%%%%%%%%%%%%%%%%%

\subsection{``GeV Excess'' Explained by Dark Matter?}

% {\bf de Boer DM model:}

In this paper we have tested the consistency of the model proposed by
\citet{deBoer} with the diffuse Galactic Plane TeV gamma-ray
observation of \citet{Milagro05}. This model requires a conventional
pionic component in order for the GeV excess to originate from WIMP
annihilation. We found that such a pionic component will then be able
to account for only $\sim 10\%$ of the Milagro TeV gamma-ray flux.
Thus, although the GeV excess could be explained this way, there still
will be a potential TeV excess.  However, due to large uncertainties
regarding point source contribution to Milagro TeV gamma-ray flux of
EGRET sources, our model-independent analysis is unable to rule
\citet{deBoer} model in or out, on the basis of gamma-rays alone. On
the other hand, the recent analysis by \citet{Bergstrom} does claim to
exclude the \citet{deBoer} model on the basis of antiproton fluxes.

\citet{Finkbeiner} has proposed that dark matter annihilations may
account for both the EGRET GeV excess and the WMAP Galactic haze,
through the inverse-Compton and synchrotron energy losses of electrons
and positrons produced in the annihilations.  Though the mechanism of
producing the GeV gamma rays is different from that of \citet{deBoer},
in both cases the gamma-ray spectrum is cut off at energies above the
dark matter mass, presumably $\sim 100$ GeV.

%%%%%%%%%%%%%%%%%%%%%%%%%%%%%%%%%%%%%%%%%%%%%%%%%

\subsection{Answering the Questions: Observations}

Some of the existing and upcoming gamma-ray experiments will be able
to answer the questions we have raised. Namely, the Gamma-ray Large
Area Space Telescope \citep[{GLAST};][]{GLAST} will make
observations in the $10 \rm \ MeV - 300 \rm \ GeV $ energy band, which
means that it will go about an order of magnitude higher in energy
than EGRET, and will thus have the first view into this ``unopened
window'' in energy.  This will give new understanding of how GeV and
TeV diffuse Galactic Plane gamma-ray observations connect. It will
tell us more about the nature of the GeV excess and how high it
extends in energy.  In particular, GLAST observations of diffuse
emission could find a break in the diffuse gamma-ray spectrum, which
would point to the nature of the GeV excess. Though a potential break
could be due to high inverse Compton component \citep{SMR}, it could
also have a dark matter origin; the shapes of two spectra may differ
enough to make separation possible.

GLAST observations of point sources at such high energies should
uncover the break in their spectra implied by the overproduction of
TeV gamma-rays when GeV data are extrapolated without a break
(Fig.~\ref{fig:sources}).  A possible break, along with in general a
better determination of point source spectra, could place strong
constraints and possibly give a definite answer about the existence of
the diffuse TeV excess.  Discovering a break in the spectra of
supernova remnants in particular would immediately have important
consequences for the nature of Galactic cosmic rays and thus hadronic
gamma-rays.  This feature would indicate a maximum cosmic ray energy
which then should also limit Galactic cosmic rays accelerated by
supernovae.  Any cosmic rays above such energies must be accelerated
from other sources, Galactic or otherwise.

The Very Energetic Radiation Imaging Telescope Array System
\citep[VERITAS;][]{veritas} will complement and partially overlap with
GLAST by observing in the energy range of 50 GeV -- 50 TeV.  VERITAS
enjoys greater flux sensitivity compared to Milagro.  Consequently,
VERITAS should better determine the intensity of diffuse Galactic
Plane gamma-ray emission.  At least as important, VERITAS has far
better point source sensitivity, which results in far lower
contamination by unresolved point sources.  All of this will allow
VERITAS to place strong constraint on the possible diffuse nature of
the TeV excess and in turn constrain the pionic gamma-ray component.

The High-Energy Stereoscopic System (HESS) is already surveying point
sources \citep{HESS}. Its sensitivity is similar to VERITAS, and thus
it is in the position to already answer some of these questions.
Although it is located in the southern hemisphere, and does not
observe the same region of the Galactic Plane as Milagro, an
independent measurement of the diffuse gamma-ray flux would help
resolve some of the issues we have presented in this paper. A possible
diffuse Galactic Plane gamma-ray measurement made by HESS could be
used to check for consistency with EGRET observations, in a similar
way as presented in this paper. The much better angular resolution of
HESS compared to Milagro would give a result far less dependent on the
unresolved point sources. Thus, a potential discovery of a diffuse TeV
excess even in this case would then tell us a lot about the nature of
this excess. We also note that the MAGIC telescope, a very large
atmospheric imaging \v{C}erenkov telescope, has a very low energy
threshold, down to 30 GeV \citep{magic}, and will thus also be a
powerful probe.

Moreover, very recently, the HESS Collaboration has reported the
discovery of an apparently diffuse flux from a very small region at
the Galactic Center \citep{HESS-Diffuse}.  While near 200 GeV, this
flux is similar to expectations, it falls off less steeply ($\alpha =
2.3$ instead of 2.75), reaching an excess of at least a factor 10 near
10 TeV.  While the spectrum here is falling less steeply than that
which would be required to explain the Milagro TeV excess, the
remarkable similarity of the excess suggests that a common origin is
possible, e.g., perhaps due to source cosmic rays \citep{Berezhko2000,
Berezhko2004}.  Note that Milagro has only measured a single point --
the flux above 3.5 TeV -- and hence cannot yet distinguish between
possible new spectra emerging near that energy.

If the enhanced gamma ray flux seen by Milagro indeed arises from
neutral pion decays, as in the model of \citet{Berezhko2000,
Berezhko2004} with enhanced high-energy cosmic ray fluxes near
sources, then it {\em must} be accompanied by an equally enhanced flux
of neutrinos from charged pion decays.  (In proton-proton collisions
at high energies, neutral and charged pions are produced in comparable
numbers; the neutral pions decay to two gamma rays, and the charged
pions ultimately decay to three neutrinos and an electron or
positron.)  The same conclusion would hold if the TeV flux excess is
due to dark matter annihilations \citep{Finkbeiner, deBoer} or
unresolved sources in which the gamma rays are produced by pion
decays.  If the excess gamma rays are produced leptonically, by
inverse Compton scattering, there will not be accompanying enhanced
neutrino fluxes.  These considerations may allow new tests of the TeV
excess in IceCube and other large neutrino detectors \citep{Beacom,
Candia, Kelley}, for which the detection prospects would be enhanced
by a factor approaching 10, and more if the excess persists to higher
energies.

Finally, as we have emphasized, gamma-ray energy spectra provide the
most direct and model-independent probe of pionic production and hence
hadronic cosmic rays.  However, the sky distribution of course also
provides important clues \citep{SMR}, and the warp in the Galactic
Plane may be helpful for separating the pionic gamma-ray component.
Since the cosmic rays are believed to be isotropic within the Galaxy,
the pionic component of the gamma ray flux should be proportional to
the column density of gas along the line of sight, whereas the inverse
Compton component depends on the radiation density.  Three-dimensional
models of the Galactic neutral hydrogen density, revealed by the
Doppler-shifted 21-cm line emission, show that the Galactic Plane is
strongly warped at radii $\ga 10$ kpc \citep{Levine}.  Some evidence
of this warp can be seen in neutral hydrogen column density maps
\citep{Kalberla}, showing up as an excursion to positive latitudes
near Galactic longitude $\ell \sim 100$ and an excursion to negative
latitudes near $\ell \sim 260$.  In the energy range corresponding to
pionic gamma rays, these same features should be seen.  While some
evidence of this effect was noted earlier \citep{Hunter}, it appears
to be easiest to see in the new EGRET maps of \citet{Cillis}, which
are shown for several energy ranges (note the high-resolution figures
are only available online).  Here the warp effect can be quite easily
seen in several of the maps, which probably implies that the
distribution of all gas is similar to that of neutral hydrogen alone.
Besides offering some hope to separate the pionic component with
spatial information, the non-trivial geometry would allow for the
first time some information about distances along the line of sight.
While our comments here are only qualitative, we are unaware of any
published correlation of the EGRET and 21-cm maps.  The future GLAST
mission, with significantly better sensitivity and angular resolution,
should allow much more detailed studies.

%%%%%%%%%%%%%%%%%%%%%%%%%%%%%%%%%%%%%%%%%%%%%%%%%
%%%%%%%%%%%%%%%%%%%%%%%%%%%%%%%%%%%%%%%%%%%%%%%%% 

\section{Conclusion:  An Observational Strategy
to Determine the Sources of Diffuse Galactic Gamma-Rays}
\label{sect:conclude}

The nature and origin of the diffuse gamma-ray emission from our
Galaxy at GeV energies \citep{Hunter} has become an increasingly
pressing problem, with the GeV excess \citep{SMR} seeming to demand
new astrophysics (e.g., high-energy cosmic-ray populations) or new
physics (annihilating dark matter).  The Milagro detection
\citep{Milagro05} of a TeV Galactic signal, possibly of diffuse
origin, invites us to place the GeV emission in a larger context.  In
this paper, we show that TeV and PeV gamma-ray observations provide a
long ``lever arm'' on the GeV excess and its origin.

In particular, the combined GeV--TeV--PeV observations shed new light
on emission due to hadronic cosmic-ray interactions.  These hadronic
gamma-rays must exist at some level, and appear with a characteristic
spectrum fixed by pion decay and the primary cosmic-ray spectrum.
Since the ``pion bump'' at $E_\gamma = m_\pi/2$ is not seen, the
evidence for pionic emission must come from the high-energy tail,
which should dominate over any leptonic (i.e., inverse Compton) signal
at high energies ($\gtrsim 1 {\rm \ TeV}$).  For this reason, TeV--PeV
data offer key new constraints on pionic gamma-rays, which can allow
us to determine the hadronic gamma component and thus isolate the
residual contribution(s).

Can we arrive at a consistent picture of high-energy Galactic
gamma-ray emission?  Yes, though present data are insufficient to
single out a unique combination of sources.  However, some conclusions
are already clear: (a) The simplest picture, in which pions are
created from cosmic-rays with energy spectra as measured locally, is
{\em inconsistent} with published EGRET and Milagro data.  (b) Besides
the ``GeV excess'' identified by EGRET, a ``TeV excess'' is likely to
be present as well. We have shown that one of the main uncertainties
in accounting for the \citet{Milagro05} diffuse TeV gamma-ray
observation comes from unresolved sources.  (c) As we have pointed
out, indications of a possible break in spectra of some point sources
can have important consequences for cosmic-ray acceleration.  (d) The
true picture of Galactic gamma-rays, which might follow several
scenarios, can be revealed with further observations.

\begin{figure}[t]
\epsfig{file=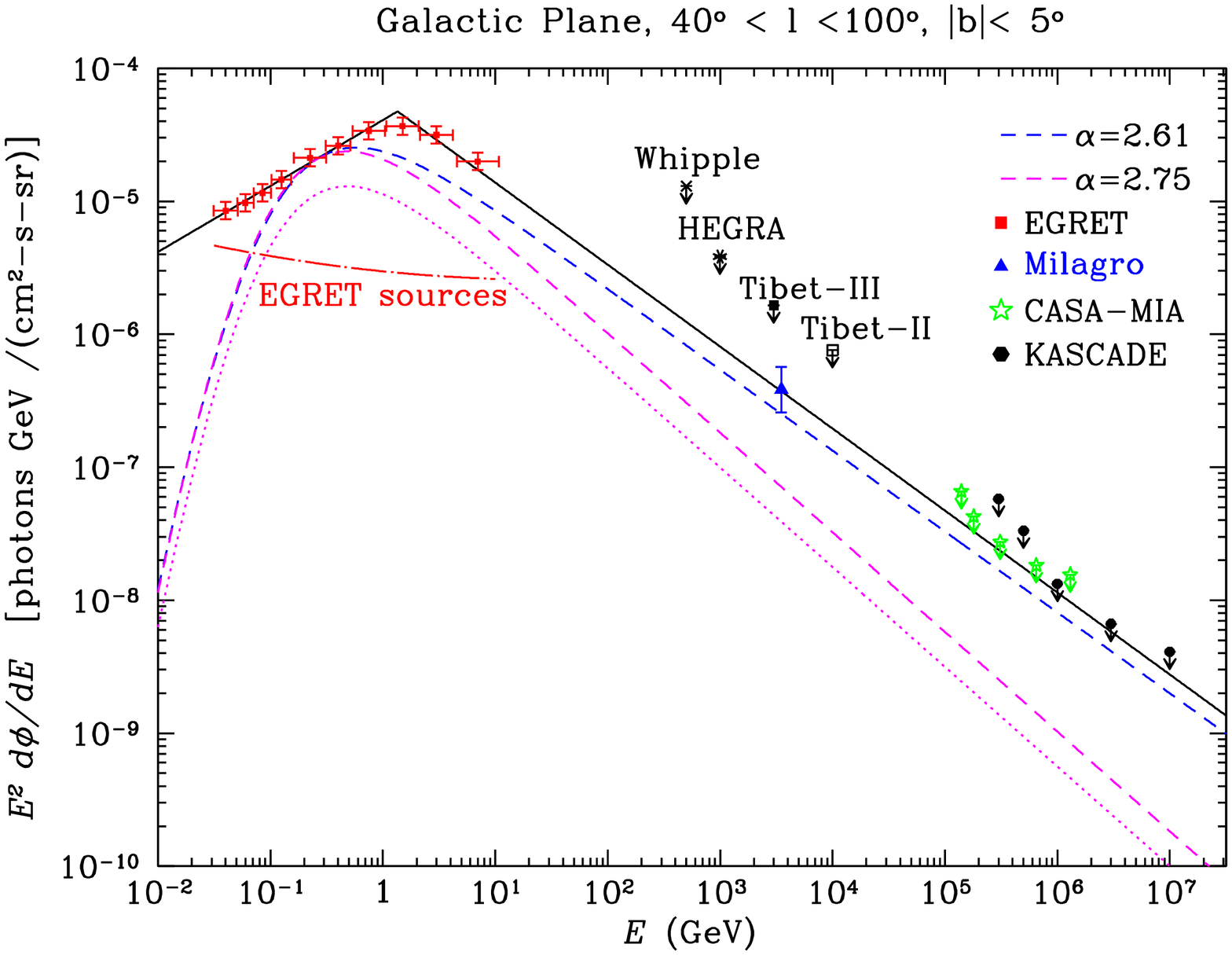, angle=-90, width=\linewidth}
\caption{The diffuse gamma-ray GeV-TeV-PeV spectrum of the Galactic
plane in the region visible to Milagro.  The EGRET data points and the
Milagro signal are empirically well-fit (solid line) with a spectral
index $\alpha=2.61$. The {\em maximized} pionic spectrum appears in
the dashed lines; we see that pionic emission having the empirical
$\alpha = 2.61$ index (dark blue line) signal comes close to (but
somewhat undershoots) the Milagro result; on the other hand, the
maximal pionic signal generated by cosmic rays with the locally
measured $\alpha=2.75$ spectrum (magenta lines) falls far short of
Milagro, leaving a TeV excess.  The dotted line represents a pionic
spectrum normalized to the one plotted in \citet{deBoer} (their
Fig.~5, region B) at $E=1 \rm \ GeV$. The PeV limits of CASA-MIA and
KASCADE are on the verge of being constraining (see also
Fig.~\ref{fig:zoom}).  Finally, fluxes of the ten EGRET sources that
we have identified were smoothed over the Milagro field of view and
then summed, which is plotted with red dash-dotted line; the Milagro
result falls below the extrapolation of this trend and thus demands a
significant break in some or all of the EGRET source spectra (see also
Fig.~\ref{fig:sources}).}
\label{fig:all}
\end{figure}

\begin{figure}[t]
\epsfig{file=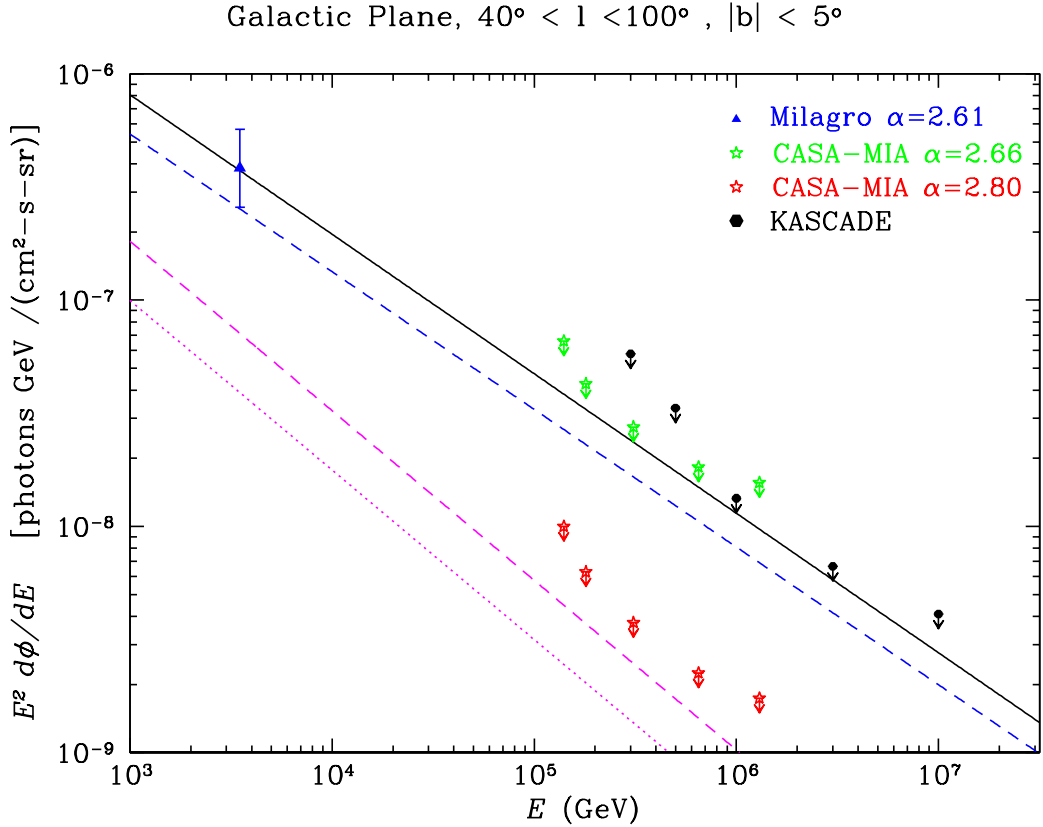, angle=-90, width=\linewidth}
\caption{In this figure, we zoom in the PeV region of
Fig.~\ref{fig:all} to emphasize the strong dependence of the CASA-MIA
limits on the assumed spectral index.  The value adopted in this paper
$\alpha=2.66$ \citep{CASA-MIA_spec} results in limits plotted as green
stars. On the other hand, if a steeper spectrum is assumed
$\alpha=2.80$ \citep[i.e., adopting the JACEE cosmic-ray
flux;][]{JACEE} this results in stronger CASA-MIA limits plotted as
red stars. As in Fig.~1, dashed lines represent the maximal pionic
spectrum for $\alpha=2.61$ (blue) and $\alpha=2.75$ (magenta), while
the pionic spectrum adopted by \citet{deBoer} is presented as a dotted
magenta line.}
\label{fig:zoom}
\end{figure}

\begin{figure}[t]
\epsfig{file=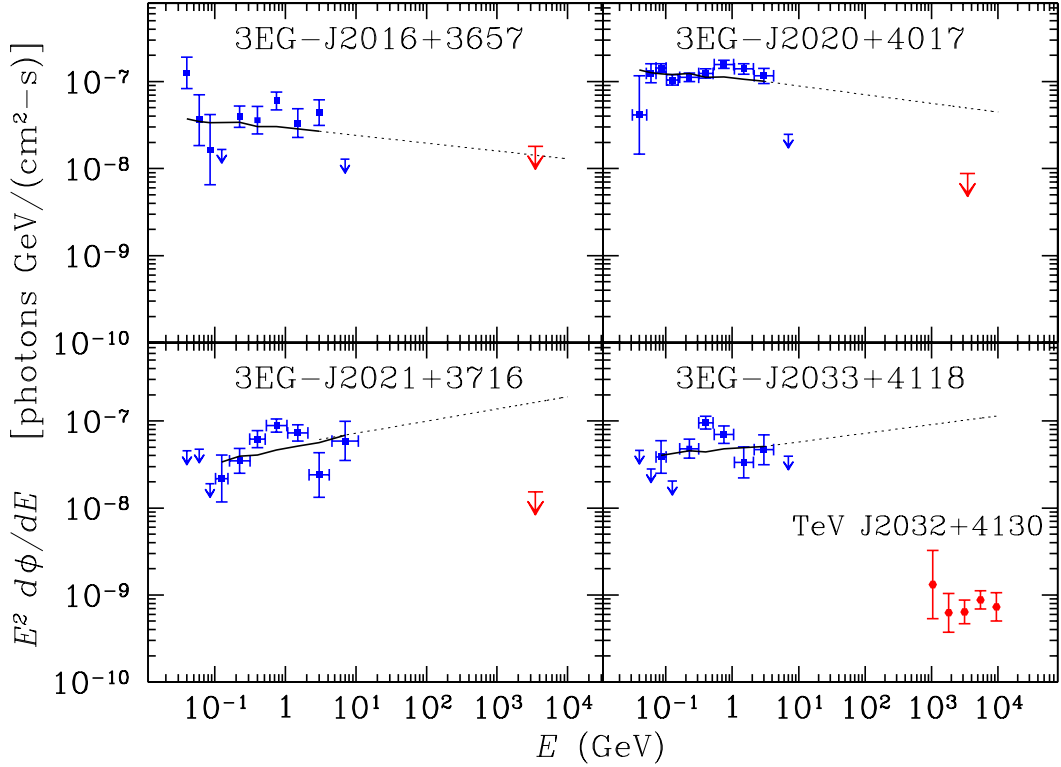, angle=-90, width=\linewidth}
\caption{In this figure we plot four of the EGRET sources from the
Milagro $\ell \in (40^{\circ}, 100^{\circ})$, $|b|<5^{\circ}$ region
that were observed in TeV range as well; we see here (but also from
Fig. \ref{fig:all}) that the power-law
trends at GeV energies must not continue to TeV energies.  The EGRET
data points (blue) were plotted using publicly available data. The
extrapolation slope used for each source is given in Table 2
\citep{Hartman}. Limits at $E=3.5 \ \rm TeV$ plotted in red were
derived from observations: J2016+3657 and J2021+3716 Whipple
\citep{Fegan05}, J2020+4017 Whipple \citep{Buckley}, J2033+4118 HEGRA
\citep{Aharonian05}.  }
\label{fig:sources}
\end{figure}

\begin{enumerate}

\item
One possibility is that the TeV excess is indeed is truly diffuse and
due to pionic emission \citep{Milagro05}.  In the simplest picture,
this would be a scenario where no break in the diffuse gamma-ray
spectrum is observed between the GeV and TeV regimes.  This would in
turn require a spectral index $\alpha=2.61$, as pointed out by the
Milagro Collaboration, which would indicate that measured local
cosmic-ray spectrum is different from, and harder than, the Galactic
average.  This would also mean that the pionic component is very close
to maximal, if not larger, as shown in Fig.~\ref{fig:all}.  
In this case, the spectrum should follow the same power law out to the
PeV region.  This PeV signal would lie just below the current limits,
awaiting discovery (or falsification!) with modest improvements in sensitivity.

Such a hard pionic spectrum would greatly reduce the GeV excess,
lessening the motivation for a large inverse Compton or dark matter
component.  For a more realistic pionic spectrum, there is the
well-known problem of the GeV excess.  We are noting here that models
which explain the GeV excess with a low pionic normalization and new
component at GeV energies must now be confronted with the TeV excess
that they create.

\item
Another possibility is that the TeV excess is truly diffuse, but not
due to interstellar pionic emission.  This would be the case if there
is a ``hard electron component,'' i.e., with a spectrum not observed
locally \citep[see e.g.,][]{aa,pe}.  Such an anomalous component could
create an inverse Compton signal which composes the GeV excess, but
also extends to TeV energies where it dominates over the pionic
component. This scenario would result in a gamma-ray spectral break at
a few tens of GeV.  Having a definite handle on the inverse Compton
component would in turn determine the pionic gamma-ray
component. Moreover, because the hard electron spectrum model explains
a large fraction of the GeV excess, it thus excludes dark matter
explanation.  However, if the TeV excess cannot be explained with the
inverse Compton component, then this would indicate a more exotic solution.

\item
Just as the GeV excess raises the exciting prospect of a dark matter
signal \citep{deBoer}, so too does the TeV excess.  This scenario is
testable.  If the TeV excess is due to annihilations, one expects a
strict cutoff above the mass of the dark matter particle (which
necessarily must be rather heavy, $m_{\rm DM} \ga E_{\gamma,\rm
Milagro} \ge 3.5$ TeV); this should appear as a break or perhaps even a
peak in the gamma-ray spectrum.  Also, the evidence for the TeV excess
comes from the Milagro region which lies $\ell > 40^\circ$ from the
Galactic center, and thus probes rather peripheral Galactocentric
radii $R > R_\odot \sin \ell \simeq 5$ kpc.  Given that dark matter
densities (and the resulting annihilation rate $\propto n_{\rm
DM}^2$) are expected to peak at the center, one would expect a rapid
increase in the diffuse signal as one scans from the Milagro region to
the Galactic center.  And a dark matter interpretation of either or
both gamma-ray signals faces similar challenges from other high-energy
particle observations \citep[e.g.,][]{Bergstrom}.

\item
Finally, the TeV excess could result from unresolved isolated sources
such as supernova remnants.  This scenario could easily be checked by
surveying for TeV point sources in the Galactic region observed by
Milagro.  Indeed, the EGRET sources in the Milagro region appear as
``hot spots'' on the Milagro map, though it is unclear how significant
this may be.  Also, another observation of the diffuse Galactic Plane
TeV gamma-rays could yield a flux that does not require a TeV excess,
but is instead consistent with a diffuse pionic emission with a more
conventional spectral index.  Thus it is crucial that VERITAS TeV
telescope surveys the EGRET sources, especially the ones in the region
observed by Milagro.

\end{enumerate}

A measurement of the diffuse Galactic Plane TeV gamma-ray
flux with better resolution telescopes like VERITAS and HESS would not
only confirm the Milagro detection, but also would provide much
sharper angular resolution of the signal.  These additional data could
significantly tighten the constraints based on gamma-ray spectra, and
open up the possibility of distinguishing the diffuse TeV sources
based on the sky distribution.

Thus, existing diffuse gamma-ray observations of the Galactic plane
are consistent with an energy spectrum that is at once empirically
simple (a broken power law) yet stubbornly resistant to theoretical
explanation.  Fortunately, upcoming observations across the
GeV-TeV-PeV range will add qualitative and quantitative detail that
will distinguish among and/or exclude the possible sources of the
highest energy photons in our Galaxy.

%%%%%%%%%%%%%%%%%%%%%%%%%%%%%%%%%%%%%%%%%%%%%%%%%
%%%%%%%%%%%%%%%%%%%%%%%%%%%%%%%%%%%%%%%%%%%%%%%%%

\acknowledgments

We thank Roman Fleysher and Peter Nemethy for discussions and for
useful comments on an earlier draft of this paper, and Frank
Fe{\ss}ler, Gavin Rowell and Gerd Schatz for helpful information. We
are also grateful to Terri Brandt, Julian Candia, Charles Dermer,
David Hanna, Evan Levine, Vasiliki Pavlidou, Christoph Pfrommer, and
Andy Strong for illuminating discussions.  The work of TP and BDF
was supported by the National Science Foundation under CAREER Grant
No.~AST-0092939.  The work of JFB was supported by the National
Science Foundation under CAREER Grant No.~PHY-0547102, and by The Ohio
State University.

%%%%%%%%%%%%%%%%%%%%%%%%%%%%%%%%%%%%%%%%%%%%%%%%%
%%%%%%%%%%%%%%%%%%%%%%%%%%%%%%%%%%%%%%%%%%%%%%%%%

%%%%%%%%%%%%%%%%%%%%%%%%%%%%%%%%%%%%%%%%%%%%%%%%%
%%%%%%%%%%%%%%%%%%%%%%%%%%%%%%%%%%%%%%%%%%%%%%%%%

%%%%%%%%%%%%%%%%%%%%%%%%%%%%%%%%%%%%%%%%%%%%%%%%%
%%%%%%%%%%%%%%%%%%%%%%%%%%%%%%%%%%%%%%%%%%%%%%%%%

% \newpage

%%%%%%%%%%%%%%%%%%%%%%%%%%%%%%%%%%%%%%%%%%%%%%%%%
%%%%%%%%%%%%%%%%%%%%%%%%%%%%%%%%%%%%%%%%%%%%%%%%%

\end{document}